\documentclass[12pt]{article}

\usepackage{color,epsfig}
\definecolor{violet}{rgb}{0.4,0,0.6}
\definecolor{vert}{rgb}{0,0.4,0.0}

\begin{document}

\def\spose#1{\hbox to 0pt{#1\hss}}\def\lta{\mathrel{\spose{\lower 3pt\hbox
{$\mathchar"218$}}\raise 2.0pt\hbox{$\mathchar"13C$}}}  \def\gta{\mathrel
{\spose{\lower 3pt\hbox{$\mathchar"218$}}\raise 2.0pt\hbox{$\mathchar"13E$}}}

\def\mm{{\color{red}m}} 
\def\ee{{\color{red}e}}
\def\ii{{\color{red}i}}
\def\rr{{\color{red}r}}

\def\ff{{\color{red}f}}
\def\alp{{\color{black}\sigma}} 

\def\tt{{\color{blue}t}}  
\def\ttau{{\color{blue}\tau}}
\def\llambda{{\color{blue}\lambda}}

\def\PP{{\color{blue} {\cal P}} }
\def\nn{{\color{red}n}} 

\def\be{\begin{equation}}
\def\fe{\end{equation}}

\font\impa = cmssbx10 scaled \magstep3
\font\impb = cmss12
\font\impc = cmss12 scaled \magstep1

\centerline {\color{red} {\impa  Five or six step scenario for evolution?} }
\vskip 1.2 cm

\centerline{ {\bf \color{violet}   Brandon Carter} }
\vskip 0.8 cm
\centerline{\color{vert} LUTh, Observatoire de Paris.}
\vskip 0.6 cm
\centerline{November, 2007.}
\vskip 1.6 cm

\noindent

{\bf{\color{blue} Abstract.}  The prediction that (due to the 
limited amount of hydrogen available as fuel in the Sun) the future 
duration of our favourable terrestrial environment will be short 
(compared with the present age of the Earth) has been interpreted 
as evidence for a hard step scenario. This means that some of the 
essential steps (such as the development of eukaryotes) in the 
evolution process leading to the ultimate emergence of 
intelligent life would have been hard, in the sense of being 
against the odds in the available time, so that they are 
unlikely to have been achieved in most of the earth-like planets 
that may one day be discovered in nearby extra-solar systems. 
It was originally estimated that only one or two of the essential 
evolutionary steps had to have been hard in this sense, but it 
has become apparent that this figure may need upward revision,
because recent studies of climatic instability suggest that 
the possible future duration of our biologically favourable 
environment may be shorter than had been supposed, only about one 
Giga year rather than five. On the basis of the statistical  
requirement of roughly equal spacing between hard steps, it is 
argued that the best fit with the fossil record is now obtainable 
by postulating the number of hard steps to be five, if our
evolution was exclusively terrestrial,  or six if, as now
seems very plausible, the first step occurred on Mars.}

\vfill\eject

\noindent
{\bf 1. \color{blue} Introduction} 
\medskip

It is generally recognised that the Darwinian process leading to 
the evolution of what we recognise as intelligent life must have 
passed through a number of essential steps, starting of course
 with the origin -- called biogenisis -- of life itself in the form
of self reproducing organisms. Another obviously important step, at a 
much later stage, is what might be called combigenesis, meaning the 
origin of sexual recombination, whereby the reproduction of genes 
ceases to be simply amalgamated with reproduction of the host organisms, 
so that evolution (in large populations) can procede much faster. What 
opinions differ about, however,  is the extent to which such  essential 
steps were easy, in the sense of being destined to occur automatically, 
given a favourable planetary environment. The aim of the present 
discussion is to clarify the problem of identifying which of the 
essential steps may have been hard in the sense \cite{Cart83} of 
depending on the fortuitous occurrence of some combination of random 
events that would automatically happen sooner or later if unlimited 
time were available, but that would be improbable within the time 
actually available.

According to the line of opinion that Davis~\cite{Dav03} has referred to as 
hypothesis B, the emergence of even the most primitive life would (due to 
the intricacy and complexity of  biological mechanisms involved) have depended 
on transitions that were hard in this sense. According to the alternative
hypothesis A,  primitive life will emerge (and perhaps be detectable 
\cite{Leger96} on extrasolar planets) by spontaneous generation or perhaps 
by panspermia wherever possible. However holders of this latter opinion are 
still divided about what follows. According to what is classifiable as 
hypothesis A-minus, after the easy establishment of primitive life, one or 
several hard steps must be achieved before the possible emergence of intelligent 
life, which will thus be very rare, even where conditions are favourable. On
the contrary, according to the more extreme alternative opinion classifiable 
as  hypothesis A-plus, not just primitive life, but even intelligent life, 
will occur (and perpaps be detectable \cite{Tart01} by the SETI program) 
wherever possible.

It was pointed out a quarter of a century ago \cite{Cart83} that evidence 
against the last of these three alternatives, hypothesis A-plus (and thus 
against the likelihood of success for the SETI program) is provided by the 
astrophysical consideration that the possible future duration of the 
favourable terrestrial environment provided by our host star, the Sun, is 
comparatively limited. The underlying reason for this limitation is that the
hydrogen still available for thermonuclear burning is sufficient for a time 
estimated to be only of the same order as the time that has already elapsed 
since the Earth was formed a little less that 5 Gyr ago. The severity of 
this already highly significant limitation has been reinforced by more 
recent work \cite{Cald92} according to which -- due to destabilisation of 
the climate by the rise in stellar temperature in the later part of the 
hydrogen burning phase -- the environmentally favourable period still 
available is reduced to the order of perhaps only 1 Gyr.

The narrowness of the margin by which we emerged on Earth so near the end of 
the time window of biological opportunity was puzzling on the basis of the 
traditional way of thinking about our Darwinian evolution just as a causal 
process within the limited framework of our own past planetary environment. 
However it can be given a reasonable interpretation -- as evidence for a hard
step scenario \cite{Cart83} -- within the broader framework invoked by the 
anthropic principle, according to which we should think of ourselves as a 
randomly selected sample within the category of comparable intelligent  
observers at other places and other times in the history of the universe.
 
The defining feature of a hardstep scenario is that one or more of the 
essential steps (such as combigenisis) in the chain leading to the evolution 
of intelligent observers is hard in the sense (as recalled above) of being 
against the odds within the allowed time. (For example, with an ordinary dice, 
getting two successive sixes would be easy if hundreds of throws were allowed, 
but if there were time for only a dozen throws it would count as a hard step.)  
Hard step scenarios can be compatible with opinions of the types listed above 
as hypothesis B or hypothesis A-minus, but evidently not with hypothesis A-plus. 
The purpose of the present article is to update the evaluation of the number of  
essential steps in our evolution that would have been hard in this sense, and to 
consider what those hards steps may have been, giving particular attention to
the question of whether they could have included  biogenisis itself, as
hypothesis B would have it.

\bigskip

\noindent
{\bf 2. \color{blue} Two step versions of the  hard step scenario}
\medskip

In simple hard step models, according to the mathematical analysis 
recapitulated in the next section, the expected interval between
the time of completion of the chain of hard steps and the end of the 
time available has the same magnitude as the expected time interval
between the hard steps, of which the last is presumably identifiable 
as the development of of the large brain needed for intelligent 
observation. On the basis of this equal spacing property, when the 
use of such a hard step scenario was originally suggested, the 
supposition  that the remaining available time interval is 
comparable to the age of the Earth implied \cite{Cart83,Mad84} that 
the total number of hard steps would only have been  one or two. 
Of these, the other earlier one -- if any -- then seemed to be 
plausibly identifiable with biogenisis itself.

With respect to the equal spacing property, the identification of 
biogenisis as the first of just two hard steps would have made sense 
if (as was supposed when its name was chosen) the onset of the 
Proterozoic eon -- when the age of the Earth was a little over 2 Gyr  
--  really had been the time of biogenisis.  However the (unexpected) 
discovery \cite{Schopf93} of what are apparently (though not quite 
certainly \cite{Bras04}) the remains of  photosynthesising bacteria 
from long before the beginning of the so called Proterozoic, can be 
considered \cite{Catl05} as rather strong evidence against this 
particular kind of two step scenario. 

A two step scenario of a more viable kind can however be obtained on 
the supposition that the first of the two hard steps was the emergence 
of eukaryotic organisms (with cell nuclei) at a time than now seems 
to fit reasonably well with the beginning of the Proterozoic, when the 
Earth was about half its present age. This revised two step scenario is 
incompatible with hypothesis B, but it is consistent with hypothesis 
A-minus, which means that it would be favoured if future 
observations~\cite{Leger96,Arnold02,Kiang07} of extrasolar
planets reveal widespread presence of primitive photosynthesizing
life systems.

The information available at present would however appear to be weighted 
(albeit not overwhelmingly) against any scenario with only two hardsteps, 
because of the increasing (but not yet absolutely conclusive) amount of 
evidence\cite{Cald92} to the effect that as remarked above, the 
environmentally favourable period still available may only be of the order 
of 1 Gyr, not of 5 Gyr as originally supposed, so that (as was suspected 
\cite{Cart83,BaTi86} from the outset) the likely number $\nn$ of hard steps 
is correspondingly larger than one or two, most probably in the range  
$4 \, \lta \nn\, \lta \, 8\, .$

\bigskip

\noindent
{\bf 3.  \color{blue} Mathematical statistics of hard step scenarios}
\medskip

The basic principle of a hardstep scenario is that, within the relevant 
environmentally favorable timescale, $\ttau_{\! \rm e}$ say, a number, $\nn$ 
say, of essential but random processes in the evolutionary chain leading to 
the outcome in question (for our purpose that of intelligent life) are hard 
in the sense of having random occurrence rates $\llambda_\ii$ ($\ii=1, ... , 
\nn$) so low that the corresponding characteristic timescales 
$\ttau_\ii=1/\llambda_\ii$ are long compared with what is available, $\ttau_\ii\gg 
\ttau_{\! \rm e}\, .$ This means that unlike other essential but easy steps, 
such hard steps will in most cases never be achieved at all, with the 
implication that the outcome in question will be rare, even in favourable
 environments (something that may become observationally verifiable when 
capabilities for observation~\cite{Leger96,Arnold02,Kiang07} of extra-solar
planets are sufficiently improved).

In a hardstep scenario of the kind  specified in this way, the (very small) 
probability, $\PP$ say, of ever completing the evolutionary chain -- leading 
in the case of interest to the emergence of intelligent observers at a 
particular site -- will be given as a product of contributions from the 
$\, \nn$ steps of the chain by
\be \PP\propto \prod_\ii \PP_\ii\, ,\hskip 1 cm \PP_\ii
=\frac{\ttau_{\! \rm e}}{\ttau_\ii}\, \ll \,1\, ,\label{steps}\fe 
while the chance of completing the chain within some given time 
$\tt\,$ (which must necessarily be less than the maximum available 
time $\ttau_{\rm e}$ will be  given by  
$\PP\{ \tt \} \propto \tt^\nn \, \Pi_\ii \llambda_\ii \, ,$  
with an order of unity proportionality factor whose exact numerical 
value depends on whether or not the steps have to be taken in a 
particular order. Independently of that, and independently of the 
values of the long timescales $\ttau_\ii\, ,$ the expected arrival time 
$\overline\tt$ (in the small fraction of cases for which the chain is 
completed) will be given by 
\be \frac{\overline\tt}{\ttau_{\! \rm e}}= \frac{\nn}{\nn+1}\, .\label{arr}\fe

On the basis of the plausible assumption that the hard steps actually 
do have to be carried out in a well defined order, it can easily be
seen that, subject to the restriction that the chain be completed 
within the allowed interval $ \ttau_{\rm e}\, ,$ the conditional 
probability for the time $\tt\{\rr\}$ of occurrence of the the $\rr$ th 
step will have a distribution, 
 $\dot P\{\rr\}= {\rm d} P/{\rm d}\tt\{\rr\}\, ,$
given by
\be \dot P\{\rr\} =\frac{\nn!\  \tt^{\rr-1}(\tt-\ttau_{\! \rm e})^{\, \nn-\rr}}
{ (\rr-1)!\ (\nn-\rr)!\  \ttau_{\! \rm e}^{\, \nn}} \, ,\label{distr}\fe
as shown, for the case $\nn=6\, ,$ in Figure 1.
It is evident that the maximum of the distribution for the $\rr$th hard 
step will occur when $\tt/\ttau_{\rm e}=(\rr-1)/(\nn-1)\, .$ This means 
that the maxima are uniformly spaced, all with with the same separation 
$\ttau_{\rm e}/(\nn -1)\, .$ For practical purposes it is more 
important to know the the corresponding mean expectation values 
$  \overline{\tt} \{\rr\}$  which  are given by the formula
\be \frac{ \overline{\tt} \{\rr\}}{\ttau_{\! \rm e}}= \frac{\rr}{\nn+1}
\, ,\label{arth}\fe
(of which (\ref{arr}) is the special case for $\rr=\nn$) from which it 
can be seen that (like the maxima) these averaged times of occurrence 
will also be evenly spaced, with separation 
\be  \Delta\overline \tt=\frac{\ttau_{\! \rm e}}{\nn+1} \, .\label{dels}\fe

Although it is highly simplified, this kind of hardstep description
is rather robust. One might seek higher accuracy by allowing for 
time variation of the rates $\llambda_\ii$, but as these  rates cancel out 
in the observationally relevant formula (\ref{arth}), and as the 
random scatter is characterised by standard deviations of at 
least  the same order as the mean separation (\ref{dels}), the 
statistical significance of improvement obtainable by such 
elaboration would hardly be enough to be worth the trouble.

When the hard step picture encapsulated in (\ref{steps}) and (\ref{arr})  
was originally put forward \cite{Cart83}, its implementation was based 
on the identification of $\ttau_{\rm e}$ (the duration of the window 
of biological opportunity) with the theoretically predicted main sequence 
(hydrogen burning) lifetime $\ttau_\odot$ of our Sun, which is of the order 
of 10 Gyr, as well as on the identification of $\overline\tt$ with the 
present age of the Earth, which is nearly 5 Gyr. The revised implementation 
here will be based on the attribution of a shorter value, only 
about 6 Gyr, to $\ttau_{\! \rm e}$, in accordance with the estimate \cite{Cald92} 
that we have already used up about five sixths of the originally available 
time before the aging Sun makes the Earth too hot. On this revised basis it 
can be seen that reasonable conformity with the formula (\ref{arr}) is 
obtained by supposing $\nn$ to be in the range 
$4\, \lta \, \nn\, \lta \, 8 \, ,$  with the best fit given perhaps by $\nn=6$.

\bigskip 
\noindent
{\bf 4. \color{blue} The six step scenario}
\medskip

If as before, one starts  by supposing that the first hard step
is biogenenis itself (including the origin of the genetic code)
then, as the final step will in any case be our own recent emergence
as very large brained animals, it remains to identify just 
4 other intermediate hard steps if we wish to complete a scenario
in which the total hard step number is $\nn=6\, .$

In view of the lack of precision of the estimate\cite{Cald92}
for $\ttau_{\rm e}\, ,$ as well as the statistical scatter of
the distributions (\ref{distr}), whose standard deviations are
at least of the same order as the mean separation (\ref{dels}), 
the optimisation of the matching of the formula (\ref{arr})
within the range $4\, \lta \, \nn\, \lta \, 8\, ,$ should not 
in itself be taken too seriously.

It has however been pointed out by Hansen \cite{Hans98} that if we 
want to match not just the final arrival formula (\ref{arr}) but 
also the formula (\ref{arth}) for the evenly distributed expected 
time of completion of the intermediate steps, then -- according to 
the new interpretation advocated by Schopf \cite{Schopf99} -- the 
fossil record provides supplementary evidence in favour of a 
scenario with just 4 intermediate hard steps, and therfore with 
total number $\nn=6\, .$  Subsequent to a first step consisting of 
biogenisis at a date too early to be evaluated today, Schopf  
identifies four successive transitions that are undoubtedly of 
cardinal importance, and that are plausible candidates for the 
status of steps that are hard in the technical sense used here, 
meaning that their occurrence within the available time 
$\ttau_{\rm e} \, \approx\,  6 \, $ Gyr was against the odds a priori. 
These steps are separated by time intervals that fluctuate from 
about 0.6 Gyr to about 1.3 Gyr, with a mean interval
 $\Delta\overline \tt$ of about  0.8 Gyr.

\begin{figure}
\centering
\epsfig{figure=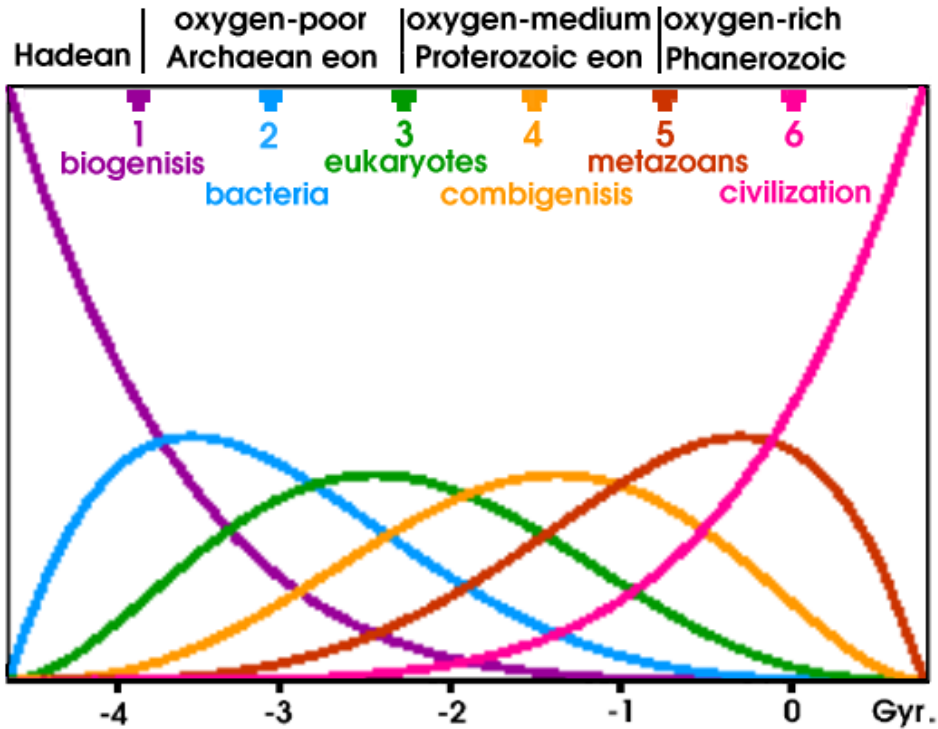, width=13.2 cm}
\caption{ Conditional probability distributions with
corresponding (numbered) expectation values and 
suggested interpretations, for a chain with $\nn=6$ hard steps
within an allowed time range that (in the chronological
scale underneath) has been taken to be nearly 6 Giga years,
so as to get the best fit to our own terrestrial case.
}
\end{figure}

The 4 intermediate steps of the Schopf list are as follows. To start 
with, the candidate for the status of the 2nd hard step is the 
emergence of procaryote (simple celled) cyanobacteria about  3.5 Gyr 
ago; the candidate for the status of the 3rd hard step is the 
emergence of eukaryotes (with cell nuclei) which were certainly 
present 1.8 Gyr ago, and for which there is evidence \cite{Brocks99} 
dating back to late Archaean times, roughly 2.5 Gyr ago; the candidate 
for the status of the 4th hard step is what I call combigenisis, meaning 
the introduction of sexual gene propagation, about 1.2 Gyr ago; and 
finally the candidate for the status of the 5th hard step is what might 
be called macromorphogenesis, meaning the emergence of metazoans (large 
multicellular animals) about 0.6 Gyr ago. On this basis, the 
emergence of our own anthropic civilization now would count as the 
sixth hard step. 

\bigskip
\noindent
{\bf 5. \color{blue} Hardsteps as transitions between eons}
\medskip

The description of the geological history of the Earth is facilitated 
by its convenient step like structure, characterised by comparatively
rapid transitions between periods during which conditions were fairly
stable, with a hierarchical structure whereby periods are grouped 
into longer units known as eras, and these are grouped into the longest 
units of all which are known as eons. The classification used in 
Darwin's time recognised only two eons: the recent relatively short
Phanerozoic eon, to which the entire macroscopic fossil record is 
limited, and the enormous pre-Cambrian super eon, which included 
everything older than about half a Gyr, but about which very little 
was known until relatively recently.

In the more modern classification commonly used today, the 4 Gyr 
pre-Cambrian super-eon has been subdivided into 3 parts. This makes
a total of 4 eons, which group into two pairs each comprising
about half of terrestrial history. It used to be thought that
life was present only in the second half, in which the
Phanerozoic eon was preceded by the much longer Proterozoic eon, 
during which only relatively simple, mainly single celled,
organisms were present. The first half started with the relatively 
brief the Hadean eon, during which conditions are thought to have 
been too extreme for survival of any life on Earth. This was
followed by the much longer and more favorable Archean eon, which 
was originally thought to have been sterile, but during which it is
now thought \cite{Schopf93} that the Earth was host to a thriving 
population of photosynthesizing cyanobacteria. It now seems 
reasonable to associate the transition from the Archaean to the
Proterozoic era with the development of eukaryotic life, in 
which the cells have an elaborate structure with chromosomes
contained in nuclei.

The recognition of these 4 rather clearly distinct eons might be 
considered as prima facie evidence in favour of a hardstep model with 
$\nn=4\, . $  However such an interpretation is disfavoured by the 
observation that the durations of these eons differ considerably, 
whereas it is to be recalled that the hardstep model predicts that the 
durations will on average be equal, with deviations that will not 
be very large compared with their mean. The fact that two of the eons 
--  namely the Archaean and  the Proterozoic -- have roughly double 
the length of the other two suggests that if the short eons -- namely 
the Hadean and the Phanerozoic -- are each associated with a single 
hard step, then the long eons should each be associated with a pair 
of hard steps, so that one finally obtains a total of 6 hard steps, 
as proposed in the preceding section, see Figure 1.

\bigskip
\noindent
{\bf 6. \color{blue} Oxygen: a convenient biproduct of combigenesis}
\medskip

A crucial issue in the interpretation of the fossil record concerns
the question (raised by Darwin himself) of why the penultimate step,
namely the emergence of metazoans, occurred at such a relatively late
time. In reply to this question, one of the key points emphasized by
Schopf and many others \cite{Catl05} is that large multicellular
organisms need an oxygen rich environment such was not available
on Earth until about the last Gyr. It has been suggested that this
 requirement should be interpreted as an astrophysical restriction,
reducing the past time duration of what should be considered as an
anthropically favourable environment from nearly 5 to less than 1 Gyr.
Taken by  itself \cite{Liv90} this interpretation would have reduced 
the estimated value of $\nn$ to zero (with the implication \cite{Liv99} 
that intelligent life could be very common) but in conjunction 
with the future limitation  \cite{Cald92} of the same order, namely 
about 1 Gyr,  it would mean simply that $\ttau_{\rm e}$ should
be interpreted as having a smaller value, of order $\ttau_\odot/5\, ,$
which would merely restore the original \cite{Cart83} estimate
$1\leq \nn\lta 2$.

It is however rather difficult \cite{Catl05} to explain the 
-- comparatively recent -- time of oxygen enrichment of the atmosphere 
on an essentially astrophysical basis. A more plausible alternative is 
to follow Schopf \cite{Schopf99} in construing the oxygen enrichment as 
part of the biological evolution of the environment. Postulating  the 
oxygen enhancement to actually be itself -- or to be an immediate 
consequence of -- one of the hard  steps in the chain suffices to restore
 the viability of the picture proposed above, in which the total available 
time, $\ttau_{\rm e}\, ,$ is taken to be between 5 and 6 Gyr, and the 
average time  $\Delta\overline \tt$ between steps is given by the estimated 
time \cite{Cald92} remaining available in the future, which is of the order 
of 1 Gyr, with the implication that the hardstep number $\nn$ is likely to 
be in the range $4 \, \lta \nn\, \lta 8\, $ which includes the particular 
suggested value $\nn=6$.

The doctrine advocated by the Schopf school is effectively as follows. 
It has long been consensually accepted that during most of terrestrial 
history the source of atmospheric oxygen (originally at a level far too 
dilute for metazoans) has been photosynthesis by  the cyanobacteria whose 
emergence is one of the most obvious hardstep candidates \cite{BaTi86}, 
counting as 2nd in the chain of 6 steps listed above, and as the first 
of the pair of hardsteps to be associated with the long Archaean eon 
(the other -- signalling the completion of the Archaean -- being the 
arrival of the eukaryotes).

The ensuing concentration of oxygen would have depended on the balance 
of this  photosynthetic production against oxygen absorbtion by various 
sink mechanisms (including combination with iron during the Archaean 
eon, prior to what is listed above as the 3rd step) of which it seems 
likely that the most important was -- and remains -- combination 
with carbon to form carbon dioxide and carbonates such as chalk.
According to an interpretation of the kind proposed by 
Schopf\cite{Schopf99} the emergence of successively more advanced life 
forms would have increased the effectiveness of inhumation processes 
whereby some of the carbon was taken out of atmospheric circulation in 
unoxidised form. The most important example of this in recent 
terrestrial history  is the conversion of buried vegetable residues to 
coal. 

Schopf has suggested that the augmentation of the proportion of 
oxygen to carbon dioxide in the atmosphere by such inhumation processes 
would have become particularly important as a convenient biproduct of 
combigenisis (the development of sex), counted as the 4th in the chain 
of 6 steps listed above, and as the first of the pair of hardsteps to
 be associated with the long  Proterozoic eon (the other -- signalling 
the completion of the Proterozoic -- being the arrival of the metazoans).
The efficient propagation of genetical material made possible by 
this innovation would (as described eldewhere \cite{Cart83})  have 
greatly increased the potental rapidity of evolution, thereby enabling  
occupation of new ecological niches by many specialised life forms of 
unprecedented diversity. The presumption is that these would have 
included kinds whose life style would posthumously produce substantial 
carbon inhumation and ensuing oil production. 

It is to be remarked that an inconvenient \cite{Gore06} biproduct of 
the rise of civilisation, counted as the 6th step in the chain, is the 
reversal of this process, by conversion of coal and oil back to carbon 
dioxide. 

\bigskip
\noindent
{\bf 7. \color{blue} The puzzle of the first hard step}
\medskip

An important question in this more definitive implementation of the hard 
step picture, as in its original application \cite{Cart83}, is
whether the first difficult step was the original development -- 
presumably by  establishing the genetic code -- of the most primitive 
forms of what we recognise as life itself. However, according to 
(\ref{arth}) as remarked above \cite{Hans98}, it is a generic feature 
of hard step scenarios that the intervals between the various 
hard steps can all be expected to have the same order of magnitude, 
$\Delta\overline \tt\, ,$ meaning, in this case, a substantial fraction 
of a Gyr  (the remaining time available in the future). On the basis of 
this consideration, the increasing amount of evidence \cite{Line02} 
suggesting that the time gap between the establishment of favorable 
conditions and the appearance of primitive life on Earth may have been 
much shorter that one Gyr has been interpreted \cite{Lineweaver02}
as implying that this was not a hard step, but should be counted as 
easy, with the implication that life (but not intelligent life) in the 
universe may be fairly common. In the five step scenario obtained in
this way, the Hadean eon would not be counted as part of the environmentally
favourable window, so the picture in Figure 1 would have to be trunkated
by removal of the first zone on the left.

Although it seems compelling at first sight, the conclusion that the 
emergence of primitive life should be relegated to the status of an easy 
step has recently been shown to be on a shakier footing than at first 
appeared. It has been pointed out by Davies \cite{Dav03} that there are 
strong reasons for believing that the relevant arena consists not of the 
single planet Earth, but of the neighbouring pair constituted by Earth 
with Mars. The idea is that primitive life in the solar system emerged 
first on Mars, where conditions would have been more favourable during 
an initial Hadean period lasting a substantial fraction of a Gyr  -- in 
other words long enough to be comparable with the  average hard step 
separation $\Delta\overline\tt\approx 0.8$ Gyr. It would have been only 
toward the end of this Martian phase -- about the beginning of the 
Archaean eon -- that conditions would have become relatively favourable 
on Earth, to which primitive life could have been transfered quite rapidly 
via meteorites. According to this rather plausible picture, the transfer 
would have counted as an easy step (due to the high rates of asteroid 
collisions at that early epoch) but the origin of the primitive life itself 
(like that of the oxygen photosynthesizers and carbon buriers later on) 
could indeed have been one -- presumably the first -- of the  hard steps, 
in which case (as supposed by hypothesis B) all kinds of life (not just 
intelligent life) in the universe would be very rare.

\vfill\eject
\bigskip
\noindent
{\bf 8. \color{blue} Conclusion: six hard steps or only five?}
\medskip

The claim\cite{Cald92} that the remaining time before destabilisation 
of the terrestrial climate by the aging Sun is only about 1 Gyr favours 
a six step or five step scenario, but if it were found to be 2 Gyr or 
more then a two step alternative would be a better bet. Although 
significant, such evidence by itself can not be overwhelming, as the 
corresponding probability distributions (see Figure 1) are rather broad 
(with standard deviations of half a Gyr or more for a six step scenario 
and three times larger for a two step scenario).  However further evidence 
reinforcing the hypothesis of a six or five step scenario (and thus 
tending to confirm the 1 Gyr estimate for the remaining available time) 
is provided \cite{Hans98} by the fossil record, in which it transpires that 
the transitions between geological eons match reasonably well with estimated 
times of occurrence of hard step candidates.

In the most plausible variant of the two step scenario, the first hard step 
does not occur until after the installation of photosynthesising bacteria, 
which would therefore occur commonly at favourable sites in extrasolar 
planetary systems, where their effects could~\cite{Leger96,Arnold02,Kiang07} 
become observationally detectable. Such a detection might provide a rather 
decisive falsification of the six step and five step scenarios, but the 
latter are for the time being what seems to be most likely on the basis of 
the limited evidence already available.  

A more delicate question is the distinction between the six step scenario
whose viability depends on the interpretation \cite{Dav03} of the Hadean 
eon as a Martian phase, and the trunkated five step scenario in which 
the Hadean is excluded from consideration as part of the environmentally 
favourable time window. This is an issue that might be settled by future 
exploration of Mars, but that would be difficult to resolve just by 
observation of extrasolar planets. The difficulty is that whereas, 
according to the five step (which in our case means exclusively terrestrial) 
scenario, the occurrence of very primitive life would have been widespread, 
its presence on extrasolar planets would probably have been ephemeral 
(depending on non renewable resources) and would usually not have engendered 
a signature of the easily detectable kind provided, in the two step scenario, 
by more advanced photosynthesizing life forms.

\end{document}